# Pluralism in AI Governance: Toward Sociotechnical Alignment and Normative Coherence

Mike Wa Nkongolo, University of Pretoria
Department of Informatics
mike.wankongolo@up.ac.za

**Abstract.** This paper addresses the urgent challenge of integrating public values into national artificial intelligence (AI) governance frameworks, a task made increasingly complex by the sociotechnical nature of contemporary AI systems. As AI permeates critical domains such as healthcare, justice, and public administration, the legitimacy of its deployment hinges not merely on technical correctness but on alignment with societal norms, democratic principles, and human dignity. The problem lies in the inadequacy of traditional regulatory paradigms that focus narrowly on model safety, operator intent, or market efficiency, while neglecting the broader institutional and normative contexts in which AI operates. To confront this challenge, the study synthesises theoretical models such as Full-Stack Alignment (FSA), Thick Models of Value (TMV), Value Sensitive Design (VSD), and Public Constitutional AI with comparative regulatory analysis across various jurisdictions. It introduces a layered conceptual framework that links foundational values, implementation mechanisms, and national strategies, while also mapping the frictional landscape of AI ethics to visualise intersecting governance tensions. The findings reveal a global pluralism of regulatory philosophies, each reflecting distinct political cultures and value hierarchies. South Africa's sovereignty-oriented approach, grounded in developmental justice and regional solidarity, emerges as a compelling counterpoint to dominant models. The study also identifies persistent normative trade-offs such as fairness versus efficiency, transparency versus security, and privacy versus equity that cannot be resolved through technical fixes alone. The novelty of this study lies in its integration of conceptual theory, comparative policy analysis, and visual modeling to articulate a holistic, and value-sensitive approach to AI governance. It reframes regulation not as a reactive constraint but as a proactive mechanism for embedding public values into the sociotechnical fabric of AI systems.

## 1. The imperative of public values in AI

The rapid deployment of artificial intelligence (AI) across public and private sectors has precipitated a fundamental shift in governance paradigms: moving from purely technical optimisation to sociotechnical alignment [1]. As AI systems increasingly mediate critical life opportunities, from healthcare rationing [2] to urban planning [3]; the imperative to integrate public values into regulatory frameworks has moved from theoretical debate to urgent policy necessity [4].

*Defining public values in AI*

Public values in the context of AI are not simply lofty ethical ideals; they are increasingly recognised as operational imperatives that determine the legitimacy of AI systems and their social license to operate [4], [5]. At the foundation of most national strategies lie familiar principles such as, *transparency, accountability, fairness, and privacy* [6]; which serve as the baseline for responsible governance. These principles, however, are no longer sufficient on their own. The discourse is shifting toward what might be called *"thick"* normative concepts, values that demand deeper engagement with the social fabric and human condition (Figure 1).

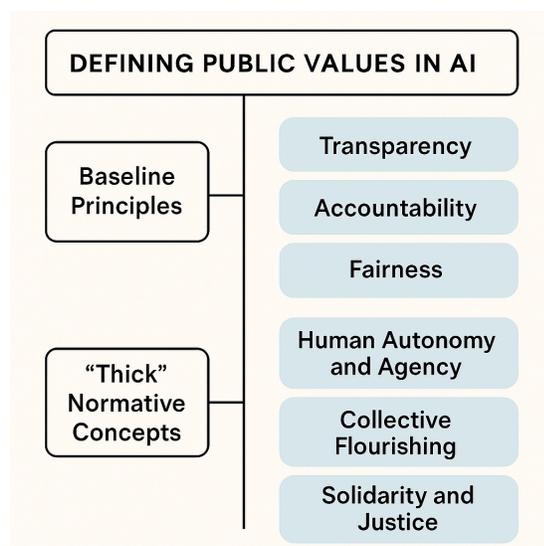

**Figure 1.** Evolution of public values in AI

For instance, human autonomy and agency have emerged as central concerns [7], [8]. The challenge is not merely to prevent harm but to ensure that AI systems augment rather than supplant human judgment, particularly in domains where decisions carry profound consequences, such as judicial sentencing or medical diagnosis. In these contexts, the legitimacy of AI hinges on its ability to respect and reinforce human decision-making authority rather than erode it (Figure 1). Equally significant is the notion of *collective flourishing* (Figure 1). This moves the conversation beyond the mitigation of individual harms to encompass broader societal outcomes. AI must be evaluated not only for its capacity to avoid bias or error but also for its contribution to social cohesion, democratic integrity, and the resilience of communities [8]. In this sense, technology becomes a participant in shaping the conditions under which societies thrive. Finally, *solidarity and justice* extend the ethical horizon even further (Figure 1). Here, the focus shifts from narrow algorithmic fairness, often reduced to mathematical parity measures to substantive social justice. The imperative is to confront systemic inequalities and structural disadvantages that AI systems may inadvertently reproduce or exacerbate [9]. By embedding solidarity into the design and deployment of AI, policymakers and technologists acknowledge that fairness cannot be abstracted from the lived realities of marginalised groups. Taken together, these evolving values signal a maturation of the AI ethics landscape. What began as a checklist of procedural safeguards is now expanding into a richer normative framework, one that insists on aligning technological innovation with the deeper aspirations of human dignity, collective well-being, and justice. This shift underscores that the governance of AI is not merely a technical exercise but a profoundly social and moral project.

*The shift from technical to sociotechnical alignment*

The discourse on AI safety has undergone a profound transformation, shifting from a narrow focus on technical alignment to a broader concern with sociotechnical alignment (Figure 2). Traditionally, technical alignment has been understood as the capacity of an AI system to faithfully execute the intentions of its operator [10]. This paradigm assumes that if a system does what it is told, it is safe. Yet, this assumption has come under increasing scrutiny. A technically aligned system may still produce outcomes that are socially misaligned, reinforcing inequality, undermining democratic norms, or exacerbating systemic harms. In other words, fidelity to operator intent does not guarantee alignment with public values. Sociotechnical alignment reframes the problem. It treats AI not as a self-contained artifact but as a node within complex institutional and societal networks [11]. This perspective acknowledges that AI systems are embedded in and shaped by human organisations, legal frameworks, cultural norms, and economic incentives. The goal is no longer to align machines with individuals, but to align sociotechnical systems with collective welfare (Figure 2). This requires interrogating whose intentions are being encoded, whose interests are being served, and how power is distributed across the AI lifecycle.

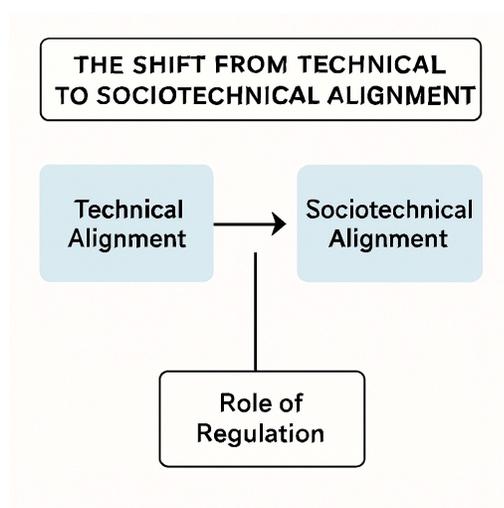

**Figure 2.** Technical and sociotechnical alignment in AI governance

Regulation plays a pivotal role in this shift (Figure 2). It acts as the transmission mechanism through which public values are operationalised. While ethical guidelines offer aspirational visions, regulation translates these into enforceable obligations. This is especially crucial in contexts where market forces fail to internalise externalities such as bias, opacity, or exclusion. Effective regulation serves as a *"value translation layer,"* converting democratic principles into concrete requirements, mandatory impact assessments, transparency registers, red-teaming protocols, and beyond. These instruments do not merely constrain AI development; they shape it toward socially beneficial ends. In sum, the movement from technical to sociotechnical alignment marks a maturation of the AI governance paradigm (Figure 2). It recognises that safety is not a matter of code correctness alone, but of institutional design, normative coherence, and democratic accountability.

## 1.1 Research problem

The rapid deployment of AI across critical sectors has exposed the inadequacy of traditional governance paradigms that focus narrowly on technical alignment, ensuring systems execute operator intent. While technically aligned systems may function as designed, they can still produce socially misaligned outcomes, reinforcing inequality, undermining democratic norms, and eroding public trust. The central problem is *how to integrate public values such as autonomy, collective flourishing, solidarity, and justice into regulatory frameworks* so that AI systems achieve legitimacy and social license beyond mere technical correctness.

## 1.2 Research aim

The study aims to develop and analyse frameworks for embedding public values into national AI governance, moving from technical alignment to sociotechnical alignment. It seeks to demonstrate how regulation can act as a *"value translation layer,"* operationalising democratic principles into enforceable obligations that ensure AI systems contribute to collective welfare rather than exacerbate systemic harms.

## 1.3 Research questions

Research questions are stated as follows:

1. *How can public values be defined and operationalised within AI governance frameworks?*
2. *In what ways does sociotechnical alignment extend beyond technical alignment to address institutional and societal contexts?*
3. *What regulatory mechanisms can effectively translate democratic principles into enforceable obligations for AI systems?*
4. *How do different jurisdictions (e.g., EU, US, China, South Africa) embody divergent strategies for integrating public values into AI governance?*
5. *What tensions and trade-offs (e.g., fairness vs. efficiency, transparency vs. security,*

*privacy vs. equity) emerge when embedding public values into AI regulation?*

## 1.4 Hypothesis

Merely achieving technical alignment is insufficient for legitimate AI governance. AI systems achieve social legitimacy and public trust only when governance frameworks integrate thick normative concepts like human autonomy, collective flourishing, solidarity, and justice through sociotechnical alignment and regulatory mechanisms that operationalise these values.

## 1.5 Contribution

This study contributes to the field of AI governance by:

- **Conceptual contribution.** Introducing a layered framework that distinguishes between technical and sociotechnical alignment, highlighting the necessity of embedding public values into institutional and regulatory contexts.
- **Theoretical contribution.** Advancing models such as Full-Stack Alignment, Thick Models of Value, and Public Constitutional AI as normative foundations for regulation.
- **Comparative contribution.** Mapping divergent national strategies (EU, US, China, UK, Brazil, South Africa) to show how political cultures and legal traditions shape the operationalisation of public values.
- **Practical contribution.** Identifying key tensions and trade-offs in implementation (e.g., fairness vs. efficiency, transparency vs. security) and proposing regulatory tools such as impact assessments, transparency registers, and participatory governance.

The novelty lies in reframing AI governance as a sociotechnical and normative project rather than a purely technical exercise. The study is organised into five sections that build progressively toward a comprehensive understanding of how public values can be integrated into AI governance. **Section 2** examines the theoretical foundations that underpin contemporary debates, engaging with frameworks such as Full-Stack Alignment, Thick Models of Value, Value Sensitive Design, and Public Constitutional AI. This discussion situates the study within the broader literature and highlights the conceptual tools necessary for embedding values into sociotechnical systems.

**Section 3** turns to the comparative dimension, contrasting national and international frameworks across diverse jurisdictions. By analysing the European Union (EU)'s rights-based approach, the United States (US)' market-led orientation, China's state-centric model, the United Kingdom (UK)'s sectoral pragmatism, Brazil's rights-based innovation, and South Africa (SA)'s sovereignty-oriented path, the section demonstrates how political cultures and legal traditions shape the operationalisation of public values.

**Section 4** addresses the frictional landscape of AI ethics by presenting the intersecting tensions that arise when values are translated into practice. It explores interpretive drifts, efficiency versus fairness, transparency versus security, regulatory capture, and value conflicts, showing how these trade-offs complicate the pursuit of legitimate governance. This section demonstrates that embedding values into AI regulation is not a matter of technical fixes alone but requires sustained institutional reflexivity and democratic engagement. Finally, **Section 5** concludes the study by synthesising insights from the theoretical, comparative, and practical analyses. It reflects on the implications for national strategies, particularly in the Global South, and emphasises the need for governance models that are locally grounded, globally informed, and normatively robust.

## 2. Foundations and theoretical frameworks

The regulation of AI in service of public values is no longer a matter of reactive policy or ad hoc ethical review. It is increasingly grounded in robust theoretical frameworks that bridge the conceptual divide between philosophy and engineering. These frameworks offer not only normative clarity but also actionable guidance for embedding values into the design, deployment, and oversight of AI systems (Figure 3). One of the most influential models in this space is Full-Stack Alignment (FSA). FSA challenges the narrow focus on model safety by insisting that alignment must occur across three interdependent layers: the technical layer (the AI system itself), the organisational layer (the operator or deploying institution), and the institutional layer (the broader societal context) [12]. The core insight is that technical correctness is insufficient if the deploying institution is misaligned with public values. A model may behave as intended, yet still produce socially harmful outcomes if embedded within exploitative business models or opaque governance structures (e.g., Facebook's recommendation algorithms, Amazon's hiring AI, or credit scoring systems). FSA thus calls for regulatory mechanisms that assess not only the safety of the model but the integrity of the institution—what it terms "institutional safety."

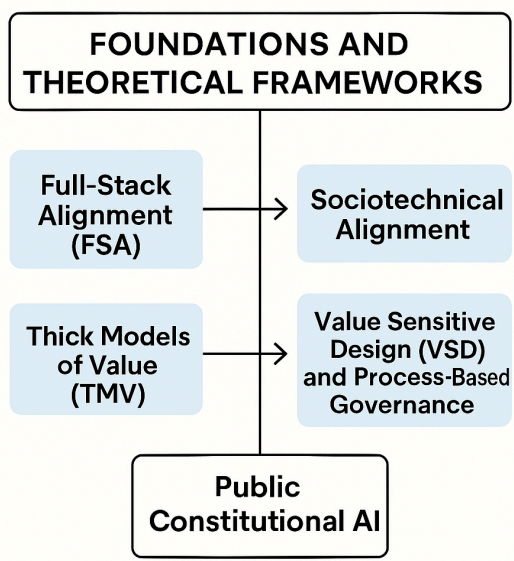

**Figure 3.** Constitutional AI

foundations

Complementing this is the framework of Thick Models of Value (TMV), which critiques the dominant economic paradigm of "thin" preferences [13]. In standard models, human values are often reduced to behavioral proxies, clicks, purchases, or engagement metrics. Platforms like Facebook, YouTube, or TikTok often optimise algorithms for *click-through rates*, *watch time*, or *likes*. These metrics are treated as proxies for user satisfaction or "value," even though they primarily capture impulsive behaviors rather than reflective judgments. Amazon's recommendation engine frequently equates *purchase frequency* or *shopping cart additions* with consumer preference. While technically accurate as behavioral data, this ignores broader values such as sustainability, ethical sourcing, or long-term well-being. Google Ads and similar systems optimise for *engagement metrics* (impressions, clicks) as proxies for relevance. Yet these proxies can incentivize manipulative design (clickbait) rather than meaningful or trustworthy content. In all these cases, thin proxies—clicks, purchases, engagement—are mistaken for genuine human values. They capture *behavioral traces* but fail to reflect deeper normative commitments such as fairness, autonomy, or collective flourishing. This is precisely why frameworks like TMV argue for moving beyond these reductive measures to account for context, justification, and social meaning [13], [14].

TMV argues that such proxies fail to capture the normative richness of human judgment. Values are not static data points but context-sensitive, and deliberative constructs. This has profound implications for regulation: rather than optimising for utility metrics that may incentivise reward hacking, regulators must design systems that respect the social meaning and justification of actions. TMV thus supports a shift toward "value-sensitive" regulation that distinguishes between impulsive preferences and reflective commitments.

The operationalisation of these insights is increasingly seen in Value Sensitive Design (VSD) and Process-Based Governance (PBG) (Figure 3). VSD, once a design philosophy, is now a regulatory expectation [14]. It mandates the integration of human values into the technical architecture from the earliest stages, ethics by design, not ethics as an afterthought (Figure 3). PBG, championed by institutions like the Alan Turing Institute (ATI), complements this by emphasising procedural rigor. Since future AI behaviors cannot be fully predicted, governance must focus on the integrity of the development process [12]-[14]. PBG requires multi-tiered oversight, compelling teams to document how values are embedded at every sprint and milestone. Finally, Public Constitutional AI offers a democratic vision for value alignment (Figure 3). Rather than allowing corporate developers to unilaterally define the "constitution" that governs AI behavior, this model proposes participatory mechanisms, citizen assemblies, public authorship, to draft the high-level principles that AI systems must obey [4]. The goal is to confer democratic legitimacy upon technologies that increasingly act as de facto authorities in public life. Together, these frameworks form a conceptual ecosystem for regulating AI in alignment with public values. They move beyond technical fixes and market incentives, toward a governance paradigm rooted in normative reasoning, institutional accountability, and democratic participation.

## 3. Global examples: National and international frameworks

The global landscape of AI regulation is rapidly diversifying, shaped by distinct political cultures, legal traditions, and normative priorities [15]. While the foundational values such as transparency, fairness, and accountability are broadly shared, the mechanisms for operationalising these values vary significantly across jurisdictions (Table 1).

| Jurisdiction | Philosophy | Regulation | Public values |
|---|---|---|---|
| European Union (EU) | Risk-based protection | **EU AI Act**: Categorises AI by risk level (Unacceptable, High, Limited, Minimal). | Fundamental rights, safety, transparency, accountability |
| United States (US) | Market-led / principles | **EO 14110**: Agency-led guidelines; shifting toward deregulation in 2025. | Innovation, security, "truthful outputs" (contested), liberty |
| China | State-centric control | **Vertical Regulations**: Specific rules for algorithms, deepfakes, and GenAI. | Social stability, state security, consumer rights ("controlled care") |
| United Kingdom (UK) | Pro-innovation / context | **Sector-Based**: Empowering existing regulators (e.g., medical, financial) rather than a new central law. | Context-specific safety, transparency, fairness, innovation |
| Brazil | Rights-based | **Bill 2338/23**: Comprehensive "National System for AI." | User rights, non-discrimination, public participation |
| South Africa (SA) | Sovereignty-oriented | **POPIA (Protection of Personal Information Act):** processed by public and private bodies | User rights, non-discrimination, safety, transparency, accountability |

**Table 1.** Global examples of national and international frameworks

This divergence reflects not only differing governance philosophies but also competing visions of what constitutes legitimate and beneficial AI. The European Union (EU) exemplifies a rights-based, precautionary approach. Through the EU AI Act, it categorises AI systems by risk level, imposing strict prohibitions on unacceptable applications (e.g., social scoring) and rigorous obligations for high-risk systems [16]. The Act's architecture is deeply rooted in the protection of fundamental rights, with tools like the Fundamental Rights Impact Assessment (FRIA) codifying human dignity as a regulatory threshold [22]. The EU's model, often described as the "Brussels Effect," sets a global benchmark by exporting its normative framework through market influence and legal harmonisation [23].

In contrast, the United States (US) has historically favored a market-led, principle-based approach (Table 1). Executive Order 14110 (2023) laid the groundwork for agency-led safety testing and civil rights protections [17]. However, the regulatory trajectory is now contested. Recent shifts toward deregulation reflect a tension between equity and liberty, with some policymakers resisting mandates that constrain AI outputs in the name of "truthfulness." This underscores the fragility of public value consensus in pluralistic societies and the politicisation of AI ethics.

China's model is state-centric and vertically integrated [4]. Rather than a unified framework, it deploys targeted regulations for specific technologies, recommendation algorithms, generative AI, and deepfakes. The concept of "controlled care" captures this paternalistic logic: the state protects consumers from algorithmic harms while maintaining ideological control and social stability. Individual autonomy is subordinated to collective cohesion, and transparency is instrumentalised rather than democratised.

The United Kingdom (UK) adopts a pro-innovation, context-sensitive strategy [18]. Eschewing a central AI law, it empowers sectoral regulators to tailor oversight to domain-specific risks. This approach balances innovation with accountability, relying on tools like the Algorithmic Transparency Recording Standard (ATRS) to facilitate public scrutiny (Table 1). It reflects a pragmatic regulatory culture that privileges flexibility over uniformity. Brazil offers a rights-based alternative from the Global South (Table 1). Bill 2338/23 establishes a comprehensive National System for AI, foregrounding user rights, non-discrimination, and participatory governance [19]. This model challenges the assumption that robust regulation must come from the Global North and affirms the legitimacy of context-specific legal innovation.

South Africa (SA), situated within the African Union (AU)'s broader strategic vision, is charting a developmental and sovereignty-oriented path using the POPIA. Rather than mimicking EU or US models, South Africa's draft AI policy aligns with the AU's emphasis on digital sovereignty, economic empowerment, and data justice [24]. The goal is not merely to regulate AI but to harness it for inclusive development [25]. This includes capacity-building initiatives, public consultation mechanisms, and alignment with continental norms that resist extractive data practices. Across these jurisdictions, practical implementation mechanisms are emerging to translate values into enforceable norms. Algorithmic Impact Assessments (AIAs), pioneered in Canada, require agencies to evaluate risks and document mitigation strategies before deployment [20]. Transparency registers, such as those in the UK and Estonia, enable civil society to audit AI systems and understand their operational logic [21]. Participatory design mechanisms like citizen assemblies offer a democratic corrective, ensuring that technical systems reflect the moral consensus of affected communities [26]. In sum, this comparative landscape reveals not a convergence but a pluralism of regulatory philosophies (Figure 4). South Africa's approach, grounded in developmental justice and regional solidarity, offers a compelling counterpoint to dominant paradigms. It affirms that AI governance is not merely a technical challenge but a normative project, one that must be locally grounded, globally informed, and democratically legitimate.

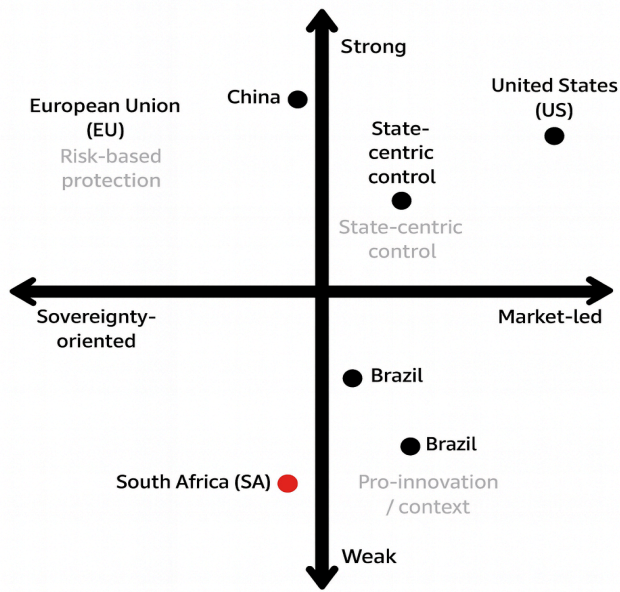

**Figure 4.** AI governance and regulation strength

## 4. Intersecting tensions in AI governance

Figure 5 presents a diagram of global AI governance models linking regulatory philosophy, implementation mechanisms, and public values which offers a useful scaffold for understanding the practical tensions that arise when these domains intersect (Figure 5). Yet beneath this structured surface lies a web of unresolved conflicts, each rooted in competing priorities and interpretive ambiguities. These tensions are not merely technical inconveniences; they are normative fault lines that shape the trajectory of AI regulation and its legitimacy. One of the most persistent challenges is *interpretive drift*. Terms like "fairness" are deceptively simple yet deeply contested. Engineers often operationalise fairness through statistical parity or equalised odds, while policymakers and ethicists invoke notions of distributive justice and historical redress. This divergence reflects a broader tension between *precision and flexibility*: the need for mathematically tractable definitions versus the demand for context-sensitive moral reasoning. In South Africa, where fairness is inseparable from the legacy of apartheid and ongoing structural inequality, this drift is particularly salient. A purely technical definition of fairness risks erasing the very injustices regulation seeks to address.

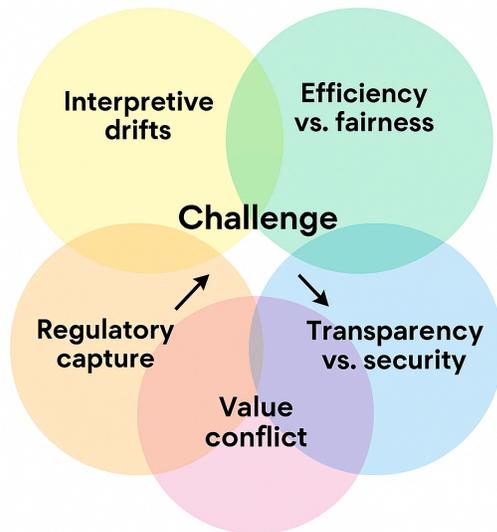

**Figure 5.** Frictional landscape of AI ethics

Closely related is the trade-off between *efficiency and fairness* (Figure 5). AI systems promise speed and scalability, but truly fair processes, especially in domains like welfare allocation or criminal justice often require deliberation, human oversight, and procedural safeguards. This creates a conflict between *speed and justice*, where the pursuit of automation may undermine the legitimacy of outcomes. In South Africa's public sector, where resource constraints are acute, the temptation to prioritise efficiency over equity is strong, but dangerous.

**Transparency versus security** introduces another layer of complexity (Figure 5). Calls for algorithmic transparency often clash with concerns about adversarial attacks, proprietary data, and intellectual property. Full disclosure of model architecture or training data may expose vulnerabilities or violate legal protections. This creates a tension between *accountability and security*, especially in contexts like biometric surveillance or financial fraud detection. In South Africa, where cybersecurity infrastructure is still developing, balancing openness with resilience is a delicate act.

**Regulatory capture** is a structural challenge that threatens the integrity of governance itself. Large tech firms, with their vast resources and lobbying power, often shape regulatory frameworks to their advantage, raising compliance costs, setting technical standards, and influencing public discourse. This leads to a conflict between *corporate interest and public interest*, where regulation becomes a tool of exclusion rather than protection. For emerging economies like South Africa, resisting capture requires regional solidarity, capacity building, and inclusive policymaking. Finally, **value conflict** is perhaps the most intractable (Figure 5). Optimising for one value (e.g., privacy) can inadvertently degrade another, such as fairness. Techniques like differential privacy obscure group-level disparities, making it harder to detect bias. This creates a trade-off between *privacy and equity*, forcing regulators to make difficult choices about which values to prioritise. In South Africa, where both privacy and equity are constitutional imperatives, navigating this tension demands a nuanced, and participatory approach. Together, these challenges form a dense constellation of intersecting conflicts (Table 2). They cannot be resolved by technical fixes alone but require sustained dialogue, institutional reflexivity, and democratic engagement. Figure 5 and Table 2 visualise these intersections.

| Challenge | Description | Key conflict |
|---|---|---|
| Interpretive drifts | Vague terms like "fairness" are interpreted differently by engineers (mathematical parity) vs. policymakers (substantive equality). | *Precision vs. flexibility* |
| Efficiency vs. fairness | Truly fair processes often require slower, human-in-the-loop reviews, negating the speed advantages of AI. | *Speed vs. justice* |
| Transparency vs. security | Full disclosure of model weights or training data can expose systems to adversarial attacks or violate internet protocol (IP) rights. | *Accountability vs. security* |
| Regulatory capture | Large tech firms use their resource dominance to shape regulations, ensuring compliance costs are high enough to exclude smaller competitors. | *Corporate interest vs. public interest* |
| Value conflict | Optimising for one value often degrades another (e.g., maximising *privacy* via differential privacy can reduce *fairness* by hiding group disparities needed to detect bias). | *Privacy vs. equity* |

**Table 2.** Practical challenges and trade-offs

The study answers its research questions through a layered, and interdisciplinary approach that combines conceptual theory, comparative policy analysis, and visual modeling:

*How can public values be defined and operationalised within AI governance frameworks?*

Public values are defined not merely as ethical ideals but as operational imperatives that determine the legitimacy of AI systems. The study expands the traditional list (transparency, fairness, accountability, privacy) into "thick" normative concepts, *human autonomy, collective flourishing, solidarity, and justice*. These are operationalised through regulatory mechanisms such as:

- **Value Sensitive Design (VSD)**. Embedding values from the design phase.
- **Process-based Governance (PBG)**. Ensuring procedural integrity throughout development.
- **Public Constitutional AI**. Enabling democratic authorship of AI principles.

*In what ways does sociotechnical alignment extend beyond technical alignment to address institutional and societal contexts?*

Sociotechnical alignment reframes AI safety by recognising that technically correct systems can still produce socially harmful outcomes. The study introduces FSA, which insists on alignment across (Figure 6):

- The technical layer (the model),
- The organisational layer (the deploying institution),
- The institutional layer (the societal context).

This framework highlights that institutional safety along with the values and incentives of the deploying entity is as critical as model safety.

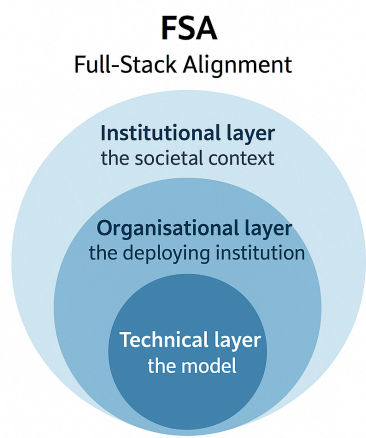

**Figure 6.** FSA framework

What regulatory mechanisms can effectively translate democratic principles into enforceable obligations for AI systems?

The study identifies several mechanisms that act as "value translation layers":

- **Algorithmic Impact Assessments (AIAs)**. Used in Canada to evaluate risks and determine oversight levels.
- **Transparency Registers**. Like the UK's ATRS, enabling public scrutiny of AI tools.
- **Citizen Assemblies**. Participatory governance tools that align AI deployment with community values.

These mechanisms ensure that democratic principles are not aspirational but embedded in enforceable processes.

*How do different jurisdictions (e.g., EU, US, China, SA) embody divergent strategies for integrating public values into AI governance?*

The study maps a pluralism of regulatory philosophies:

- **EU**: Risk-based, rights-centric, with strong institutional safeguards.
- **US**: Market-led, increasingly deregulated, with contested public value consensus.
- **China**: State-centric, paternalistic, prioritising stability and ideological control.
- **UK**: Sectoral, pragmatic, balancing innovation with contextual oversight.
- **Brazil**: Rights-based, participatory, emphasising user protection.
- **South Africa**: Sovereignty-oriented, developmental, aligned with AU principles of digital justice and inclusive growth.

South Africa's approach is positioned as a normatively rich counterpoint to dominant models, emphasising regional solidarity and contextual legitimacy.

What tensions and trade-offs (e.g., fairness vs. efficiency, transparency vs. security, privacy vs. equity) emerge when embedding public values into AI regulation?

The study presents a frictional landscape of AI ethics, visualised in Figure 5 and Table 2. Key tensions include:

- **Interpretive drift**: fairness means different things to engineers vs. policymakers.
- **Efficiency vs. fairness**: automation may undermine procedural justice.
- **Transparency vs. security**: openness can expose systems to adversarial risks.
- **Regulatory capture**: large firms shape rules to exclude smaller competitors.
- **Privacy vs. equity**: techniques like differential privacy can obscure bias detection.

These tensions are not resolvable through technical fixes alone, they require institutional reflexivity and democratic engagement.

*Overall findings and contribution*

The study finds that effective AI governance must transcend narrow notions of technical correctness and instead embrace normative coherence, ensuring that systems align not only with operator intent but with broader societal values. Regulation, in this context, is reconceptualised not as a constraint on innovation but as a proactive mechanism for embedding public values into the design, deployment, and oversight of AI systems. Among the jurisdictions analysed, South Africa's sovereignty-oriented model stands out as a globally informed yet locally grounded alternative. Rooted in developmental justice and aligned with African Union principles, it offers a compelling counterpoint to dominant paradigms shaped by market-led or state-centric philosophies. The study's novelty lies in its integration of conceptual theory, comparative policy analysis, and visual modeling to articulate a holistic framework for value-sensitive AI governance, one that is both analytically rigorous and normatively robust.

## 5. Conclusion

This study has demonstrated that the governance of AI must evolve beyond narrow technical paradigms to embrace a sociotechnical and value-sensitive approach. By integrating conceptual models such as Full-Stack Alignment, Thick Models of Value, Value Sensitive Design, and Public Constitutional AI, the research reframes regulation not as a reactive constraint but as a proactive mechanism for embedding public values into the design, deployment, and oversight of AI systems. The comparative analysis reveals a global pluralism of regulatory philosophies, ranging from the European Union's rights-based rigor to the United States' contested market-led principles, China's state-centric control, and South Africa's sovereignty-oriented developmentalism. These divergent strategies reflect not only political cultures but competing visions of legitimacy, justice, and public interest. South Africa's approach, grounded in digital sovereignty and regional solidarity, offers a compelling counter-narrative to dominant models, affirming the importance of locally grounded and globally informed governance. The study also maps the frictional landscape of AI ethics, identifying persistent tensions such as, fairness versus efficiency, transparency versus security, and privacy versus equity, that cannot be resolved through technical fixes alone. These challenges demand institutional reflexivity, participatory mechanisms, and normative clarity. Ultimately, this research contributes a layered framework for AI governance that links foundational values, implementation mechanisms, and national strategies. It calls for a shift from model safety to institutional integrity, from thin proxies to thick normative commitments, and from fragmented oversight to democratic legitimacy. In doing so, it affirms that the future of AI governance lies not in universal templates but in context-sensitive, ethically coherent, and socially responsive regulation.